\documentclass[twocolumn,english,aps,prl,floats,showpacs]{revtex4}
\usepackage[OT1]{fontenc}
\usepackage[latin1]{inputenc}
\usepackage{amsmath}
\usepackage{graphicx}
\usepackage{amssymb}

\makeatletter
\usepackage{graphicx}
\usepackage{amssymb}

\usepackage{amsfonts}
\usepackage{graphicx}

\usepackage{babel}
\makeatother
\begin{document}

\title{Einstein-Podolsky-Rosen correlations via dissociation of a molecular
 Bose-Einstein condensate}

\author{K.~V. Kheruntsyan$^{1}$, M.~K. Olsen$^{1,2}$, and P.~D. Drummond$^{1}$}

\affiliation{$^{1}$ARC Centre of Excellence for Quantum-Atom Optics,
 School of Physical Sciences, University of Queensland, Brisbane,
Qld 4072, Australia\\
 $^{2}$Instituto de F\'{\i}sica da Universidade Federal Fluminense, 
Boa Viagem 24210-340, Niter\'oi - Rio de Janeiro, Brazil}

\date{\today {}}

\begin{abstract}

Recent experimental measurements of atomic intensity correlations through atom shot noise suggest that atomic quadrature phase correlations may soon be measured with a similar precision. We propose a test of local realism with mesoscopic numbers of massive particles based on such measurements. Using dissociation of a Bose-Einstein condensate of diatomic molecules into bosonic atoms, we demonstrate that strongly entangled atomic beams may be produced which possess Einstein-Podolsky-Rosen (EPR) correlations in field quadratures, in direct analogy to the position and momentum correlations originally considered by EPR. 
\end{abstract}


\maketitle

The recent demonstrations of atomic correlation measurements at the shot noise level~\cite{Bloch,Jin} are a 
significant step towards true quantum atom optics. Quantum optics, which began with photon  correlation measurements, has
allowed for many fundamental tests of quantum mechanics.
Importantly, the availability of lasers allowed the development of techniques to perform 
quadrature phase measurements. In quantum \emph{atom} optics, Bose-Einstein condensates (BEC) play the role of the laser. However, homodyne noise 
correlation measurements of \emph{atomic} field quadratures have not been available. 

In this Letter, we suggest one route to achieve this, and 
outline a scheme which would allow for fundamental tests of quantum mechanics with massive particles. We base our proposal on the Einstein-Podolsky-Rosen (EPR)
paradox~\cite{EPR}. The EPR paper introduced two particles with perfect
correlations (entanglement) in momenta and positions,
these persisting with spatial separation.
Depending on which property of one particle we choose to measure,
we can predict with certainty the same observable of the other particle.
EPR concluded that local realism
was inconsistent with the completeness of the quantum mechanical description
of nature. As suggested by Reid in $1989$~\cite{eprMDR}, products of variances of inferred optical 
quadratures can demonstrate the paradox by seeming to violate the Heisenberg
uncertainty relation, although this is impossible for directly observed quadratures.
This is applicable to realistic correlations, and was
demonstrated experimentally by Ou \textit{et al.}~\cite{Ou} in $1992$.

We show here that dissociation~\cite{Moelmer2001,twinbeams,Dissociation-exp,Rempe-shell}
of a molecular BEC can also exhibit
EPR correlations in atomic quadratures. Such tests of quantum mechanics (see
also Ref. \cite{4WM-Meystre-Zoller}) are a step toward
understanding the properties of mesoscopic superpositions of massive
particles, since they introduce couplings to gravitational
fields, not previously tested in quantum measurement experiments. 

There has been much experimental progress~\cite{WFHRH2000,Donley-Cs-Rb-Na} and
intense theoretical interest~\cite{PDKKHH-1998,Superchemistry-KKPD-2000,Timmermans,%
Mackie-Javanainen,Yurovski-Julienne-Ben-Reuven,Verhaar,Holland,Hope-Olsen-etal,%
Goral-Rzazewski-2001, Koehler-Burnett-2003,Duine-Stoof}~%
in the production of molecular dimers
from Bose condensed atoms.
We will assume that this can be created from
bosonic constituents and propose a realization of the EPR paradox
via dissociation, which can automatically yield two counter-propagating
beams through momentum conservation. Starting from a three-dimensional
(3D) molecular condensate, D\"{u}rr \textit{et al.}~\cite{Rempe-shell}
have used Feshbach resonance techniques to create a quasi-mono-energetic
expanding wave of bosonic atoms, in close analogy to successful
fermionic correlation experiments~\cite{Jin}. 
Using well-known optical trapping techniques~\cite{Tolra}, a 1D bosonic experiment would give the two beams needed here.

We consider an initially phase-coherent atomic BEC confined to one
spatial dimension. (Ref.~\cite{1D-Bose-gas}
gives the necessary conditions.) This is then divided into three parts.
The two outside parts are moved away from the central core, and stored
for use as local oscillators. The central core is coherently converted
to a molecular BEC. Our theory describes the result of a subsequent
dissociation of the molecular BEC into two energetic 
\textquotedblleft daughter\textquotedblright\ condensates~\cite{twinbeams},
which interfere with the local oscillators to provide the measured
quadrature signals.

The quantum-field-theory effective Hamiltonian describing this process is ~\cite{twinbeams,PDKKHH-1998}
\begin{align}
\hat{H} & =\hat{H}_{0}+\hbar\int dz\left\{ \sum_{i}V_{i}(z)\hat{\Psi}_{i}^{\dag}\hat{\Psi}_{i}^{{}}+\sum\limits _{i,j}\frac{U_{ij}}{2}\hat{\Psi}_{i}^{\dag}\hat{\Psi}_{j}^{\dag}
\hat{\Psi}_{j}\hat{\Psi}_{i}\right.\notag\\
 & \left.-i\frac{\chi(t)}{2}\left[\hat{\Psi}_{2}^{\dagger}\hat{\Psi}_{1}^{2}
-\hat{\Psi}_{1}^{\dagger\,2}\hat{\Psi}_{2}\right]\right\}.
\label{eq:Ham}
\end{align}
Here, the atomic and molecular fields are respectively described by the bosonic operators $\hat{\Psi}_{1}$
and $\hat{\Psi}_{2}$, $\hat{H}_{0}$ is the kinetic energy, $V_{i}(z)$ ($i=1,2$) are the trapping
potentials (including internal energies), and the $U_{ij}$ are the
strengths of the one-dimensional intra- and cross-species $s$-wave
interactions. The term $\chi(t)=\chi_{0}\theta(t_{1}-t)$
is responsible for coherent conversion of molecules into atom pairs, where
$\chi_{0}>0$ and $\theta(t_{1}-t)$ is the Heaviside function that
turns off the coupling $\chi$ at $t>t_{1}$. 

In what follows, we assume that the atom-atom $s$-wave scattering
term is negligible. This condition is imposed in order to minimize
the phase diffusion~\cite{Steel98} of the two local oscillators required
to access the atomic quadrature correlations. This requires either
a low density and short interaction times, or else an effective renormalized
interaction near a magnetic Feshbach resonance which is tuned to give
a zero effective scattering ($U_{11}=0$). The dissociation coupling
$\chi_{0}$ would be caused in the first
case by a Feshbach sweep, or in the second case~\cite{Feshbach-plus-Raman} 
by a coherent Raman transition
with an overall detuning $2\Delta$ \cite{twinbeams}. This gives
an energy mismatch $2\hbar\Delta<0$ between the atomic and molecular
fields, which is converted into kinetic energy {[}$2\hbar|\Delta|\rightarrow2\hbar^{2}k^{2}/(2m_{1})${]}
of atom pairs with opposite momenta around $k_{0}=\pm\sqrt{2m_{1}|\Delta|/\hbar}$, 
where $m_{1}$ is the mass. 

To gain some analytic insight, we will first analyze a simple
model, beginning with a uniform molecular BEC in a coherent state
with 1D (linear) density $n_{2}$. The condensate extends from
$-L/2$ to $L/2$, with periodic boundary conditions. The dissociation
coupling $\chi$ is turned on suddenly, and subsequently assumed to
be constant. Because we are interested in evolution over short times,
the molecular field depletion is assumed at this stage to be negligible
so that the amplitude $\Psi_{2}=\sqrt{n_{2}}$ (assumed real) can
be absorbed into an effective gain constant $g=\chi_{0}\sqrt{n_{2}}$.
The dimensionless form of the equations is achieved by introducing
characteristic time and length scales, $t_{0}=1/g$ and $d_{0}=\sqrt{\hbar t_{0}/(2m_{1})}$,
and transforming to dimensionless time $\tau=t/t_{0}$, coordinate
$\xi=z/d_{0}$, detuning $\delta=\Delta t_{0}=\Delta/g$, and dimensionless
fields $\hat{\psi}_{i}(\xi,\tau)=\hat{\Psi}_{i}(\xi d_{0},\tau t_{0})/\sqrt{n_{2}}$.
The dimensionless initial molecular field density is now scaled to one. 

We expand $\hat{\psi}_{1}(\xi,\tau)$ in terms of single-mode bosonic operators:
$\hat{\psi}_{1}(\xi,\tau)=\sum_{q}\hat{a}_{q}(\tau)e^{iq\xi}/\sqrt{l}$,
where $q=d_{0}k$ is a dimensionless momentum {[}$l=L/d_{0}$, $k=(2\pi/L)n$,
$n=0,\pm1,\pm2,...${]}.
The corresponding Heisenberg equations have the solutions $\hat{a}_{q}(\tau)=A_{q}(\tau)\hat{a}_{q}(0)+B_{q}(\tau)\hat{a}_{-q}^{\dagger}(0)$,
and $\hat{a}_{-q}^{\dagger}(\tau)=B_{q}(\tau)\hat{a}_{q}(0)+A_{q}^{\ast}(\tau)\hat{a}_{-q}^{\dagger}(0)$,
where $A_{q}(\tau)=\cosh\left(g_{q}\tau\right)-i\lambda_{q}\sinh\left(g_{q}\tau\right)/g_{q}$,
$B_{q}(\tau)=\sinh\left(g_{q}\tau\right)/g_{q}$, $\lambda_{q}\equiv q^{2}+\delta$,
and $g_{q}\equiv(1-\lambda_{q}^{2})^{1/2}$. The coefficients $A_{q}$
and $B_{q}$ satisfy $|A_{q}|^{2}-B_{q}^{2}=1$. The detuning $\delta$
is the only parameter that characterizes the dynamics of this dimensionless
model. For dissociation to proceed, $\delta$ must be negative, which
can be achieved by appropriate tuning of the frequencies of the Raman
lasers. 

In the above solutions, coupling is between opposite momentum components
only. In quantum optics similar solutions have been studied by Reid~\cite{eprMDR}
in the context of parametric down-conversion. In that case, the parameter
$\lambda_{q}$ would be identified with an effective phase mismatch
term, which was set to zero. We note here that, unlike photons, the
correlated atom pairs are not distinguishable by frequency or polarization,
but by different momenta or spatial locations. We can now calculate
any operator moments at time $\tau$, given a vacuum initial state
of the atomic fields. 

We now consider the measurements that
must be made to demonstrate the EPR paradox. It is readily seen that
correlations exist between atomic quadratures of the beams with opposite
momenta. For example, a measurement of $\hat{X}_{q}=\hat{a}_{q}+\hat{a}_{q}^{\dag}$
allows us to infer, with some error, the value of $\hat{X}_{-q}=\hat{a}_{-q}+\hat{a}_{-q}^{\dag}$,
and vice versa. The same holds for the $\hat{Y}_{\pm q}=-i(\hat{a}_{\pm q}-\hat{a}_{\pm q}^{\dag})$
quadratures. This allows us to define, depending on which beam we
measure, four inferred variances, 
\begin{equation}
V^{inf}(\hat{X}_{\pm q})=V(\hat{X}_{\pm q})-[V(\hat{X}_{+q},\hat{X}_{-q})]^{2}/V(\hat{X}_{\mp q}),
\label{eq:V-inf}
\end{equation}
with similar expressions for $V^{inf}(\hat{Y}_{\pm q})$, where $V(a,b)=\langle ab\rangle-\langle a\rangle\langle b\rangle$.
These quadrature correlations can be studied using
balanced homodyne detection, which mixes the signal with a strong
local oscillator on the matter-wave analog of a 50-50 beam splitter
and is a well-known technique in quantum optics. Quadratures are measured
via measuring the density differences, after combining the signal
and local oscillator~\cite{Hansbachor}. 

As an example demonstrating the EPR paradox we consider the correlations
between the momentum components $\pm q_{0}=\pm\sqrt{|\delta|}$ corresponding
to perfect phase matching with $\lambda_{q}=0$. In this case we obtain
$V^{inf}(\hat{X}_{+q_{0}})V^{inf}(\hat{Y}_{+q_{0}})=\cosh^{-2}(2\tau)<1$,
and the same result for $V^{inf}(\hat{X}_{-q_{0}})V^{inf}(\hat{Y}_{-q_{0}})$.
Since the products of the non-inferred variances 
are bound by the Heisenberg uncertainty relation $V(\hat{X}_{\pm q_{0}})V(\hat{Y}_{\pm q_{0}})\geq1$,
this is similar to the EPR paradox. However, as the plane-wave momentum components
are not localized, this system does not allow the required spatial
separation of the EPR \emph{gedanken} experiment. 

We now return to the more realistic case described by the full Hamiltonian
(\ref{eq:Ham}). The dissociated atoms are assumed untrapped longitudinally
yet confined transversely, so that they can be treated as a free 1D
field, initially in a vacuum state. The absolute detuning $|\Delta|$
must not exceed the trap radial oscillation frequency $\omega_{\perp}$
in order to maintain the 1D condition~\cite{1D-Bose-gas} with \textquotedblleft frozen\textquotedblright\ transverse
motion of the dissociated atoms. For completeness, the atom-molecule
and molecule-molecule scattering terms are all taken into account.
Dimensionless interaction couplings are introduced according to $u_{i2}=U_{i2}\sqrt{n_{2}}/\chi_{0}$, 
where $n_{2}=n_{2}(0)$ is the initial 1D peak density
of the molecular BEC. 

To solve for the resulting quantum dynamics, we use stochastic differential
equations in the positive-P representation~\cite{+P,Steel98}. The essence of 
the positive-P method is in mapping the operator equations of motion 
into stochastic $c$-number differential equations that can be solved 
numerically. This requires four independent stochastic fields, $\psi_{i}$ and $\psi_{i}^{+}$,
corresponding to the operators $\hat{\psi}_{i}$ and $\hat{\psi}_{i}^{\dag}$,
while $v_{1}(\xi,\tau)=u_{12}\psi_{2}^{+}\psi_{2}$ and $v_{2}(\xi,\tau)=-u_{22}(1-\xi^{2}/\xi_{0}^{2})+\sum_{i}u_{i2}\psi_{i}^{+}\psi_{i}$
represent the effective potentials including the atom-molecule and
molecule-molecule $s$-wave interactions. Here, $\xi_{0}=z_{0}/d_{0}$
is the dimensionless Thomas-Fermi (TF) radius. We include linear losses
of atoms and molecules at rates $\gamma_{i}$. The stochastic variables
have a correspondence with normally ordered operator moments in the
sense of an average over a large number of trajectories. The equations
are \begin{align}
\frac{\partial\psi_{1}}{\partial\tau} & =i\frac{\partial^{2}\psi_{1}}{\partial\xi^{2}}-(\gamma_{1}+i\delta+iv_{1})\psi_{1}+\kappa\psi_{2}\psi_{1}^{+}\notag\\
 & +\sqrt{\kappa\psi_{2}}\;\eta_{1}+\sqrt{-iu_{12}\psi_{1}\psi_{2}/2}\;(\eta_{2}+i\eta_{3}),\notag\\
\frac{\partial\psi_{2}}{\partial\tau} & =\frac{i}{2}\frac{\partial^{2}\psi_{2}}{\partial\xi^{2}}-(\gamma_{2}+iv_{2})\psi_{2}-\frac{\kappa}{2}\psi_{1}^{2}\notag\\
 & +\sqrt{-iu_{22}\psi_{2}^{2}}\;\eta_{4}+\sqrt{-iu_{12}\psi_{1}\psi_{2}/2}\;(\eta_{2}-i\eta_{3}),\notag\\
\frac{\partial\psi_{1}^{+}}{\partial\tau} & =i\frac{\partial^{2}\psi_{1}^{+}}{\partial\xi^{2}}-(\gamma_{1}-i\delta-iv_{1})\psi_{1}^{+}+\kappa\psi_{2}^{+}\psi_{1}^{{}}\notag\\
 & +\sqrt{\kappa\psi_{2}^{+}}\;\eta_{5}+\sqrt{iu_{12}\psi_{1}^{+}\psi_{2}^{+}/2}\;(\eta_{6}+i\eta_{7}),\notag\\
\frac{\partial\psi_{2}^{+}}{\partial\tau} & =\frac{i}{2}\frac{\partial^{2}\psi_{2}^{+}}{\partial\xi^{2}}-(\gamma_{2}-iv_{2})\psi_{2}^{+}-\frac{\kappa}{2}\psi_{1}^{+2}\notag\\
 & +\sqrt{iu_{22}\psi_{2}^{+2}}\;\eta_{8}+\sqrt{iu_{12}\psi_{1}^{+}\psi_{2}^{+}/2}\;(\eta_{6}-i\eta_{7}),\label{Positive-P-eqs}\end{align}
Here, $\eta_{j}$ ($j=1,...8$) are real, independent Gaussian noise
terms with the correlations $\overline{\eta_{j}(\xi,\tau)\eta_{k}(\xi^{\prime},\tau^{\prime})}=\delta_{jk}\delta(\xi-\xi^{\prime})\delta(\tau-\tau^{\prime})$,
and $\overline{\eta_{j}}=0$. To numerically integrate these equations,
we consider that the molecular BEC is initially in a coherent state,
represented spatially by the TF solution. The time duration for the
molecule-atom conversion is controlled via $\kappa(\tau)=\theta(\tau_{1}-\tau)$,
so that $\kappa(\tau)=0$ for $\tau>\tau_{1}$. 
Once the dissociation stops, we continue the
evolution of the resulting atomic field in free space to allow spatial
separation of atoms with positive and negative momenta. At the same
time we set the molecular fields to zero for $\tau>\tau_{1}$. This
models selective removal of the molecules by a \textquotedblleft
blast\textquotedblright\ pulse with a resonant laser \cite{Dissociation-exp}
and is aimed at minimizing the effect of atom-molecule scattering
and atom losses due to inelastic collisions, which can potentially
reduce the atom-atom correlations. We note that the losses due to
inelastic collisions are neglected altogether in our model. With typical
loss rate coefficients of the order of $5\times10^{-17}$ m$^{3}$/s
\cite{Dissociation-exp} and peak molecular density $\sim 10^{20}$ m$^{-3}$,
their disruptive effect should be negligible on submillisecond timescales
used here. Similar considerations apply to the role of losses due
to molecule-molecule inelastic collisions and three-body losses \cite{Tolra}. 

In the nonuniform treatment, due to the mode mixing of different momentum
components, there is a difference in the way the necessary quadratures
must be defined in comparison to our analytic plane-wave treatment
(see also Ref. \cite{Yurovski-entangled}). With the quadratures defined
in terms of individual Fourier components as above, we obtain no inferred
violation of the uncertainty principle and hence no EPR correlation
signature. This is because of the assumption of uniform local oscillators,
implicitly built into this definition of the quadratures. 
This problem also arises in the measurement of pulsed optical squeezing
\cite{PDpulse} and is overcome using pulsed local oscillators~that
are mode-matched with the signals. Accordingly, we define four mode-matched
quadrature operators as \begin{align}
\hat{X}_{\pm}(\tau) & =\int d\xi\lbrack\phi_{\pm}^{\ast}(\xi)\hat{\Psi}_{1}(\xi,\tau)
+\phi_{\pm}(\xi)\hat{\Psi}_{1}^{\dagger}(\xi,\tau)],\\
\hat{Y}_{\pm}(\tau) & =-i\int d\xi\lbrack\phi_{\pm}^{\ast}(\xi)\hat{\Psi}_{1}(\xi,\tau)
-\phi_{\pm}(\xi)\hat{\Psi}_{1}^{\dagger}(\xi,\tau)].\end{align}
Here, $\phi_{\pm}(\xi)=|\phi_{\pm}(\xi)|\exp(\mp iq_{0}\xi-i\vartheta)$
are two nonuniform local oscillator fields having the same center-of-mass
momenta, $\pm q_{0}=\pm\sqrt{|\delta|}$, as the two respective atomic-beam
signals. These are described classically, with Gaussian profiles,
$|\phi_{\pm}(\xi)|^{2}=(2\pi\sigma_{\pm}^{2})^{-1/2}\exp(-\xi^{2}/2\sigma_{\pm}^{2})$,
with $\int d\xi|\phi_{\pm}(\xi)|^{2}=1$, and are centered at the
locations of the twin atomic beams at the time of measurement $\tau$.
The phase $\vartheta$ of the local oscillators can be optimized to
compensate for the molecular mean field phase shift. 

The new quadrature 
operators have the same commutation relations as
before so that the EPR criterion, 
$V^{inf}(\hat{X}_{\pm})V^{inf}(\hat{Y}_{\pm})<1$, and the expressions
for the inferred variances are the same. Experimentally, the fields
$\phi_{\pm}$ should be larger than the atomic signals, but they have
been normalized to one here for convenience. %

\begin{figure}[t]
\includegraphics[height=2.8cm]{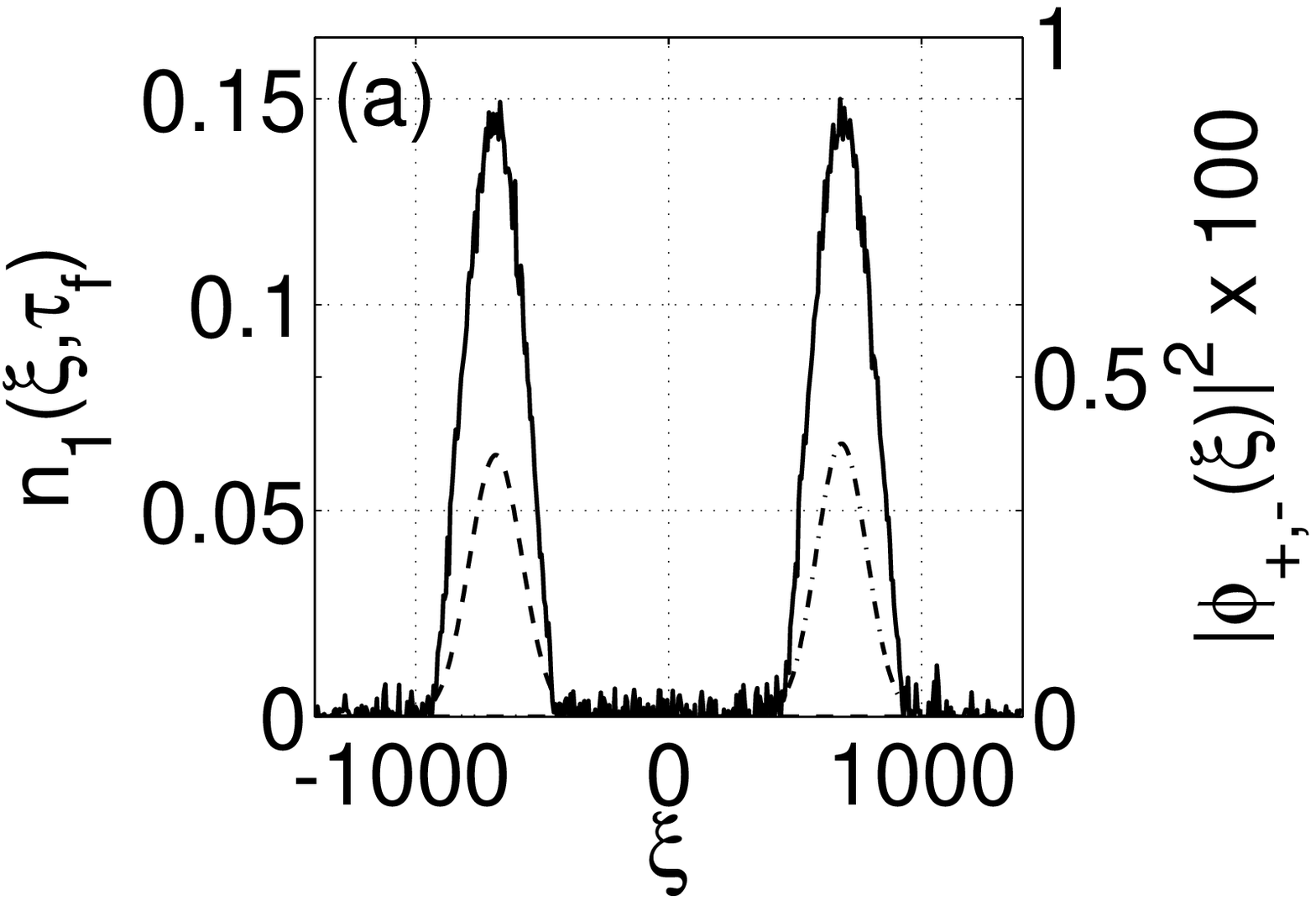}~\includegraphics[height=2.75cm]{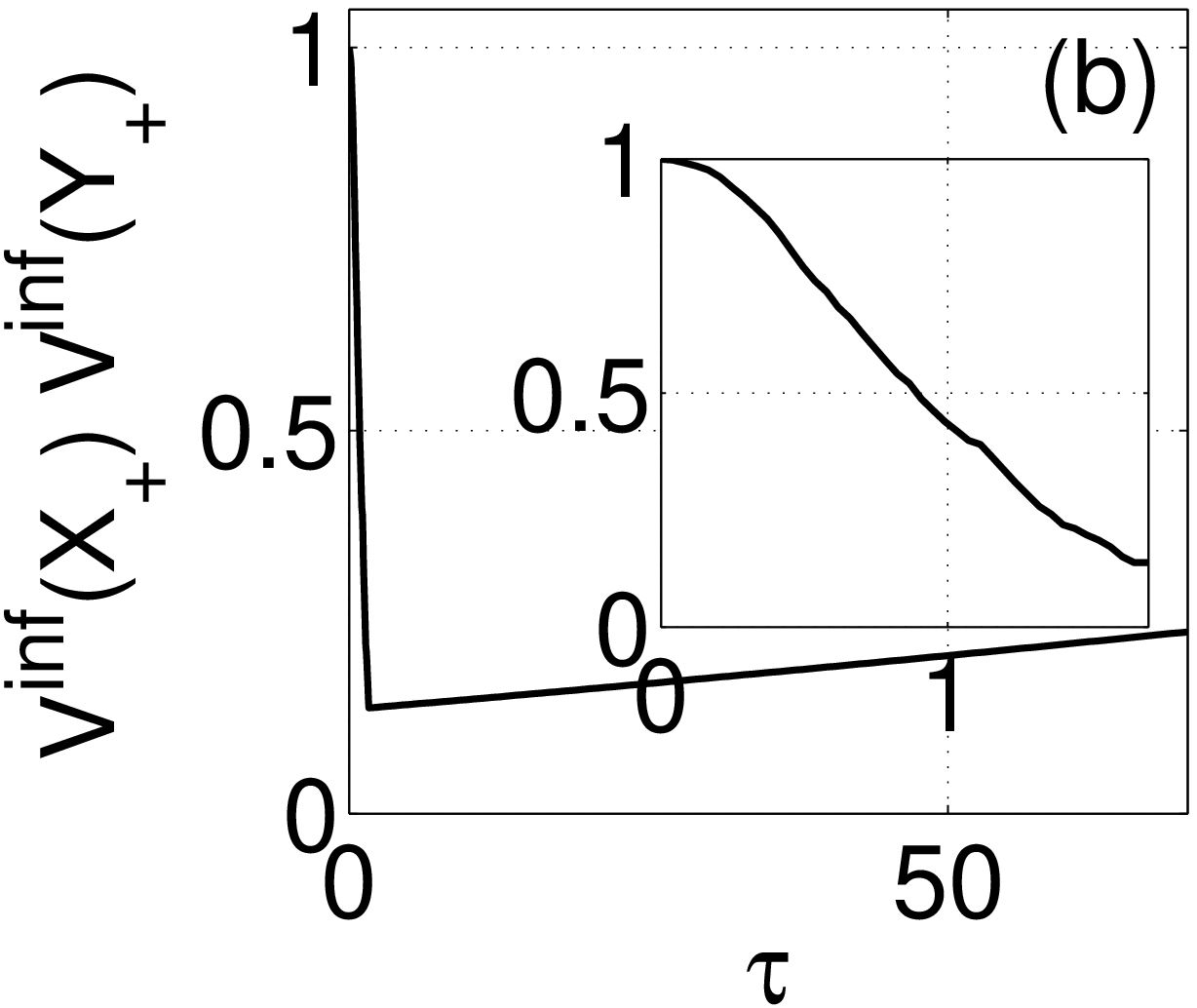}
\caption{(a) Final average atomic density distribution $n_{1}(\xi,\tau_{f})$
(solid line) and the densities of the normalized local oscillator
fields $|\phi_{+}(\xi)|^{2}$ and $|\phi_{-}(\xi)|^{2}$ (dash-dotted
and dashed lines, with $\sigma_{+}=1.45$ and $\sigma_{-}=1.50$,
respectively).
We used a time window of $\tau_{f}=70$, with dissociation on from
$\tau_{0}=0$ to $\tau_{1}=1.65$. Other parameter values are $\delta=-25$,
$u_{22}=0.2$, $u_{12}=0.1$, $\xi_{0}=238$, and 
$\gamma_{1(2)}=1.2\times10^{-3}$.
(b) Product of the resulting inferred variances as a function of time
$\tau$ for $\vartheta=0.19$. The inset shows the same quantity while
the dissociation is on.}
\label{densities}
\end{figure}

Figure~\ref{densities}a shows the final atomic and local oscillator
densities, obtained from simulation of the full Eqs. (\ref{Positive-P-eqs}),
averaged over $50,000$ stochastic trajectories. In this example \cite{parameters},
the average number of atoms in each beam after dissociation is $52$,
with about $10\%$ of them being lost during the subsequent free expansion stage.
Note that, while the two local oscillators should share the same phase,
they can have slightly different shapes, or atom numbers, without
a strong destructive effect on the correlations. As we mentioned earlier, 
the local oscillators
can in principle be prepared by splitting a single atomic BEC and
then \textquotedblleft stored\textquotedblright\ at spatial
locations away from the molecular BEC. The zero relative phase can
be maintained by Feshbach tuning of the magnetic field to
the zero crossing of the effective atom-atom scattering length. In
this case, the relative phase drift due to possible unequal number
of atoms in the two local oscillators is minimized. 

Figure~\ref{densities}b shows the product of the inferred variances,
giving a clear demonstration of the EPR paradox as 
$V^{inf}(\hat{X}_{+})V^{inf}(\hat{Y}_{+})<1$. We note that quadrature
correlations studied here require simultaneous measurement of many
particles, and that repeated measurements of single atomic pairs would
not yield the same results. 

To summarize, we have shown that dissociation of a molecular BEC into
bosonic atoms can provide a simple yet robust demonstration of the
EPR paradox with massive particles. The effects of molecular condensate
trapping and depletion, atom- and molecule-molecule $s$-wave interactions, and
possible one-body losses of atoms and molecules have all been included
in our numerical calculations. An experimental realization of our
proposal would be the first step towards testing fundamental quantum mechanics with
mesoscopic numbers of massive particles. 

The authors acknowledge support from the Australian Research Council,
the New Zealand Foundation for Research, Science and Technology (Grant
No. UFRJ0001), and thank the authors of the XMDS software~\cite{XmdS}.


\begin{thebibliography}{10}
\bibitem{parameters}Considering a $^{87}$Rb$_{2}$ experiment as a guideline, these parameters
relate to a molecular BEC confined by a harmonic trap with an axial
(radial) frequency $\omega_{z}/2\pi=1$ Hz ($\omega_{\perp}/2\pi=12$
kHz). We chose the molecule-molecule scattering length $a_{22}=1$
nm, giving the 1D coupling 
$U_{22}\simeq 2\hbar a_{22}/m_{2}l_{\perp}^{2}=1.5\times10^{-4}$
m s$^{-1}$, where $m_{2}=2m_{1}$ and $l_{\perp}=\sqrt{\hbar/m_{2}\omega_{\perp}}$.
In addition, we take the atom-molecule $s$-wave interaction strength
$U_{12}=0.75\times10^{-4}$ m s$^{-1}$; the relatively small values
of these couplings are chosen here for making the computation feasible.
The initial number of molecules is $N_{2}=380$, at 1D peak density
of $n_{2}(0)=\pi l_{\perp}^{2}n_{2}^{(3D)}(0)=3.1\times10^{6}$ m$^{-1}$
($n_{2}^{(3D)}(0)=2\times10^{20}$ m$^{-3}$), while the TF radius
is $z_{0}=\sqrt{2\hbar U_{22}n_{2}(0)/(m_{2}\omega_{z}^{2})}=93$
$\mu$m . We take the detuning $\Delta=-6.1\times10^{4}$ s$^{-1}$,
and the atom-molecule coupling $\chi^{(3D)}=3.4\times10^{-7}$ 
m$^{3/2}$ s$^{-1}$, 
so that the 1D value is $\chi\simeq\chi^{(3D)}/\sqrt{4\pi l_{\perp}^{2}}=1.4$
m$^{1/2}$ s$^{-1}$. The characteristic time scale is $t_{0}=0.41$
ms, so that the duration of dissociation ($\tau_{1}=1.65$) corresponds
to $t_{1}=0.68$ ms, while the total time window ($\tau_{f}=70$)
is $t_{f}=29$ ms.
\end{thebibliography}
\end{document}